\documentclass[manuscript,screen]{acmart}

\AtBeginDocument{%
  \providecommand\BibTeX{{%
    \normalfont B\kern-0.5em{\scshape i\kern-0.25em b}\kern-0.8em\TeX}}}

\setcopyright{acmlicensed}
\copyrightyear{2024}
\acmYear{2024}
\acmDOI{XXXXXXX.XXXXXXX}

\acmConference[ISS '24]{ACM ISS 2024}{October 27--30,
  2024}{Seattle, WA}
\acmISBN{978-1-4503-XXXX-X/18/06}

\newcommand{\cPCTwoD}{\texttt{PC2D}}
\newcommand{\cPCThreeD}{\texttt{PC3D}}
\newcommand{\cVRTwoD}{\texttt{VR2D}}
\newcommand{\cVRThreeD}{\texttt{VR3D}}

\newcommand{\cPCTwoDVRTwoD}{\cPCTwoD{}\texttt{+}\cVRTwoD{}}
\newcommand{\cPCTwoDVRThreeD}{\cPCTwoD{}\texttt{+}\cVRThreeD{}}
\newcommand{\cPCThreeDVRThreeD}{\cPCThreeD{}\texttt{+}\cVRThreeD{}}
\newcommand{\cPCThreeDVRTwoD}{\cPCThreeD{}\texttt{+}\cVRTwoD{}}

\newcommand{\cPCTwoDPCTwoD}{\cPCTwoD{}\texttt{+}\cPCTwoD{}}
\newcommand{\cVRThreeDVRThreeD}{\cVRThreeD{}\texttt{+}\cVRThreeD{}}

\newcommand{\fried}[2]{$\chi^2=#1, p=#2$}
\newcommand{\pValue}[1]{$p=#1$}

\usepackage{subfigure}
\usepackage{array}
\usepackage{booktabs, makecell, tabularx, enumitem}
\newcommand{\added}{\textcolor{black}}
\newcommand{\addedTwo}{\textcolor{black}}
\newcommand{\addedThree}{\textcolor{black}} 

\begin{document}

\title{Evaluating Layout Dimensionalities in PC+VR Asymmetric Collaborative Decision Making}

\author{Daniel Enriquez}
\email{denriquez@vt.edu}
\affiliation{%
  \institution{Virginia Tech}
  \city{Blacksburg}
  \state{Virginia}
  \country{USA}
  \postcode{24061}
}

\author{Wai Tong}
\orcid{0000-0001-9235-6095}
\affiliation{%
  \institution{Texas A\&M University}
  \city{College Station}
  \state{Texas}
  \country{USA}}
\email{wtong@tamu.edu}

\author{Chris North}
\affiliation{%
  \institution{Virginia Tech}
  \city{Blacksburg}
  \state{Virginia}
  \country{USA}
  \postcode{24061}
}
\email{north@cs.vt.edu}

\author{Huamin Qu}
\affiliation{%
  \institution{The Hong Kong University of Science and Technology}
  \city{Hong Kong}
  \country{China}}
\email{huamin@cse.ust.hk}

\author{Yalong Yang}
\orcid{0000-0001-9414-9911}
\affiliation{%
  \institution{Georgia Tech}
  \city{Atlanta}
  \state{Georgia}
  \country{USA}}
\email{yalong.yang@gatech.edu}

\renewcommand{\shortauthors}{Enriquez, et al.}

\begin{abstract}
With the commercialization of virtual/augmented reality (VR/AR) devices, there is an increasing interest in combining immersive and non-immersive devices (e.g., desktop computers) for asymmetric collaborations.
While such asymmetric settings have been examined in social platforms, significant questions around layout dimensionality in data-driven decision-making remain underexplored. 
A crucial inquiry arises: although presenting a consistent 3D virtual world on both immersive and non-immersive platforms has been a common practice in social applications, does the same guideline apply to lay out data? 
Or should data placement be optimized locally according to each device's display capacity? 
This study aims to provide empirical insights into the user experience of asymmetric collaboration in data-driven decision-making. 
We tested practical dimensionality combinations between PC and VR, resulting in three conditions: PC2D+VR2D, PC2D+VR3D, and PC3D+VR3D. 
The results revealed a preference for PC2D+VR3D, and PC2D+VR2D led to the quickest task completion. 
Our investigation facilitates an in-depth discussion of the trade-offs associated with different layout dimensionalities in asymmetric collaborations.
\end{abstract}

\ccsdesc[300]{Human-centered computing~Mixed / augmented reality}
\ccsdesc[300]{Human-centered computing~Virtual reality}
\ccsdesc[500]{Human-centered computing~Collaborative interaction}
\ccsdesc[300]{Human-centered computing~User studies}

\keywords{cross-platform collaboration, data-driven collaboration, immersive analytics, virtual reality}

\begin{teaserfigure}
  \includegraphics[width=\textwidth]{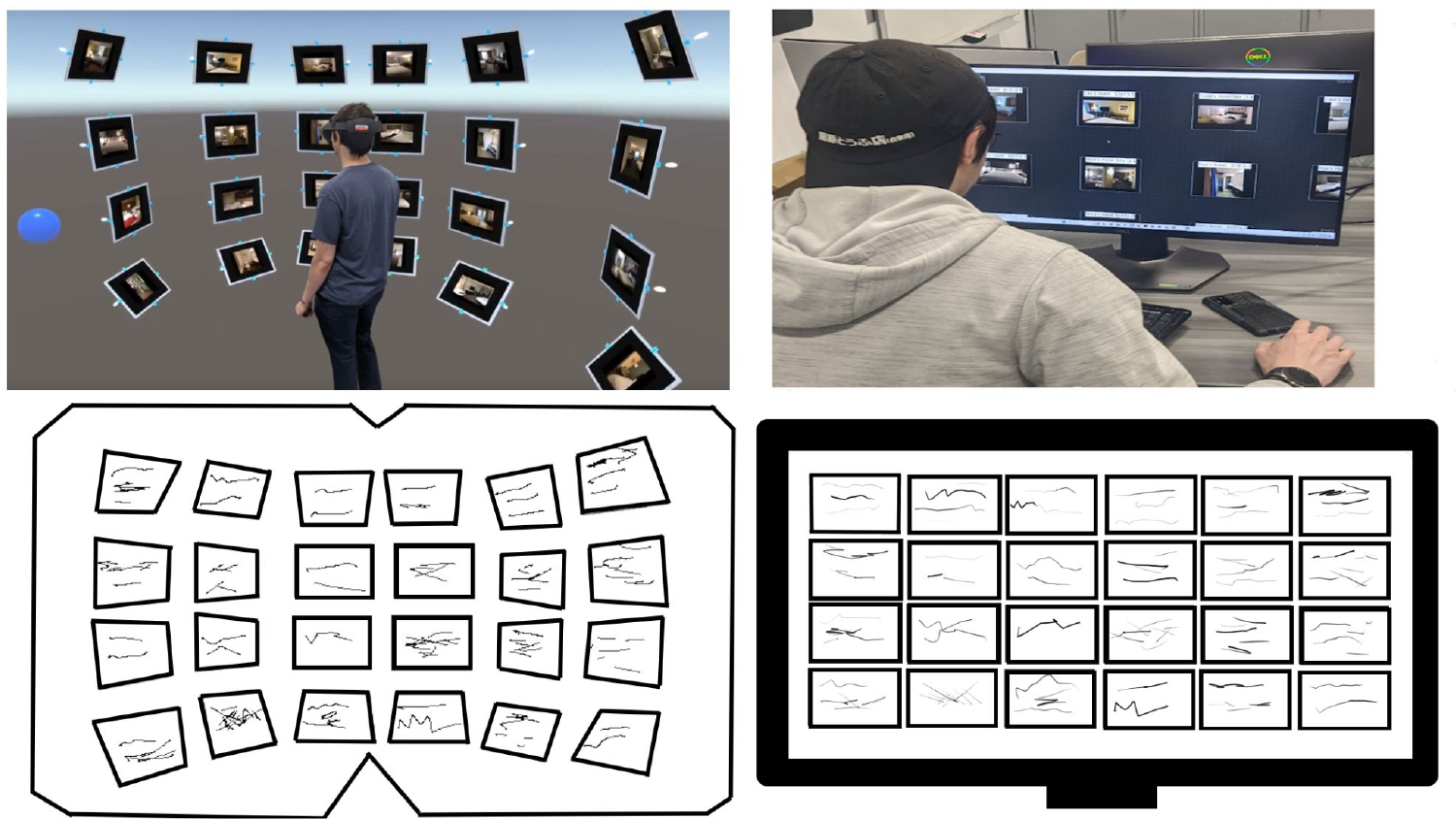}
  \caption{A VR user and a PC user are collaborating to find the best hotel that suits their needs from the same set of hotels. Showcased is the study task that was tested in the controlled study and a sketch of the view of each user. }
  \label{fig:teaser1}
\end{teaserfigure}

\received{20 February 2007}
\received[revised]{12 March 2009}
\received[accepted]{5 June 2009}


\maketitle

\section{Introduction}

Collaborative decision-making leverages diverse perspectives to improve the quality and accuracy of decisions, resulting in stakeholder satisfaction and sustainable outcomes~\cite{isenberg2011collaborative,bannon1989cscw}.
It is a demanding and challenging process that necessitates stakeholders to continuously exchange and discuss findings, formulate and evaluate hypotheses, and ultimately reach conclusions~\cite{mahyar2014supporting}.
Consequently, the exploration of technologies that can effectively support this process has been a fundamental and enduring research focus within the field of human-computer interaction.

The desktop computing environment, commonly known as Personal Computers (PCs), is the most widely utilized computing environment. As a result, extensive research has been conducted on collaborative systems involving multiple PCs~\cite{rodden1991survey}.
Today, a growing array of technologies has emerged, presenting new opportunities for beyond-desktop computing~\cite{roberts2014visualization,bellotti1996walking}.
Notably, the commercialization of virtual reality (VR) and augmented reality (AR) devices has accelerated at an astonishing pace, poised to become an integral part of our daily lives in unprecedented ways~\cite{laviola20173d,bowman20083d}.
The capacity to deliver interactive 2D and 3D graphics within VR/AR environments offers an opportunity to revolutionize our interaction with data and holds immense potential for facilitating collaboration~\cite{sereno2020collaborative,schafer2022survey}.

As the range of available computing environments continues to expand, it has become increasingly common for individuals to employ different devices to collaboratively accomplish tasks~\cite{brudy2019cross}.
\addedTwo{Our focus lies in the realm of asymmetrical collaboration across the two opposite sides of the virtuality continuum~\cite{milgram1995augmented} given that PCs maintain their position as the mainstream computing environment with limited capabilities of immersion, while VR/AR represents the emerging next-gen display for immersive interaction.}
The integration of non-immersive and immersive environments presents exciting challenges and opportunities in interaction and user interface designs.
Within PC+VR collaboration, one area of substantial interest is remote instruction, where PC users provide guidance for VR users. However, real-world decision-making processes often demand intensive two-way communication. 
To this end, various social platforms have been created, such as VirBela~\cite{Virbela}, Meta's Horizon Workrooms~\cite{MetaHorizonWorkrooms}, and games~\cite{Spatial,squadrons,nomanssky}.
While these platforms emphasize the importance of social presence, they consistently render digital content in 3D for both PC and VR. 
Nevertheless, there are additional factors that can impact the effectiveness of collaborative data-driven decision-making. 
Specifically, the collaborative process encompasses both collective and individual components~\cite{gutwin_design_1998,gutwin_descriptive_2002}. 
While maintaining a uniform presentation of information aids in the collective aspects, it may not be optimal for individual tasks. 
Furthermore, there exists an alternative method for ensuring consistency across PC and VR: in addition to employing 3D for both PC and VR, one may also consider utilizing 2D for both platforms, particularly in the context of data-driven decision-making.

In light of the aforementioned trade-offs and design choices in PC+VR asymmetric collaboration, our objective is to investigate how individuals communicate, exchange information, and engage in discussions while dealing with different combinations of dimensionalities ~\addedTwo{given the PC's inclination for 2D interactivity space and VR's 3D interactivity}.
Inspired by prior studies~\cite{in2023table,liu2020design,lisle2021sensemaking}, we study this problem within a data-driven context, wherein multiple pieces of information are displayed in various windows, requiring collaborators to arrive at a consensus and make informed decisions based on the provided data \addedTwo{ through the visualization concept of} \addedThree{grid layouts}.
In terms of the information itself, we have employed a well-established data-driven decision-making task from literature~\cite{jetter2011materializing}, where pairs of participants collaborate to select the most suitable hotel from a given set.

We have chosen three practical PC+VR designs, taking into account the dimensional aspects of PC and VR environments: \cPCTwoDVRTwoD{}, \cPCTwoDVRThreeD{}, and \cPCThreeDVRThreeD{}. 
Specifically, \cPCTwoDVRThreeD{} optimizes settings separately for PC and VR, facilitating individual work \addedTwo{ as PC is optimized for a 2D interface and VR is optimized for a 3D interface}.
Meanwhile, \cPCTwoDVRTwoD{} and \cPCThreeDVRThreeD{} provide interfaces with the same dimensionality, either in 2D or 3D, for both PC and VR, enhancing \addedTwo{collaboration given the similar interface}. 
Drawing insights from data collected from 18 pairs of participants, we have observed that optimizing user interfaces and interactions for a users' individual environment \addedTwo{(i.e., individual effectiveness)} positively influences the perceived user experience \addedTwo{resulting in participants generally preferring \cPCTwoDVRThreeD{}.}
Additionally, maintaining a consistent \addedTwo{dimensionality between both users means} resulted in shorter completion times, with \cPCTwoDVRTwoD{} emerging as the fastest overall. 
\addedTwo{Furthermore, due to the challenges posed by navigating a 3D environment on a PC, \cPCThreeDVRThreeD{} maintained a lower completion time, despite being common practice in asymmetric social applications.}

\addedThree{
This work's contribution is explored through an asymmetric collaboration system that facilitates collaboration between PC and VR environments.  
Through this, we conducted a controlled user study to investigate the trade-offs between individual and collaborative effectiveness in PC+VR collaboration, given the difference in layout dimensionalities across conditions.
}

\section{Related Work}
\subsection{Immersive Analytics}
Typically, data analysis tasks have been relegated towards PC environments; however, there is significant precedent to change this, given that immersive technologies provide tools that allow for sensemaking which provides spacial context towards thoughts that promote data interpretation~\cite{chandler2015immersive}.
As analysis tasks grow complex, other mediums should be leveraged to grow and facilitate understanding of analysis tasks~\cite{roberts2014visualization}.
Technologies such as tabletop displays allow users to understand data by providing more context for awareness~\cite{isenberg2009collaborative}. The same benefit could be obtained from immersive head-worn devices (HWDs).
Many visualizations leverage this extra dimension, and much prior work has gone into detail on the advantages and disadvantages Immersive Analytics provides ~\cite{serrano2022immersive,fonnet2019survey,kraus2022immersive}.
As a result, a few of the relevant and key works will be discussed to provide insight into the direction to lead our work rather than providing an exhaustive list of immersive analytics.

\added{Immersive analytics has a strong potential to facilitate analysis. However, less work was investigated in collaborating between immersive technology and traditional workflow with desktops. It is crucial for immersive Analytics as different users have different preferences, accessibility, and capabilities in using different devices for analytics, especially since immersive technology is relatively new and requires higher learning curves.}
Specifically, Ens et al.~\cite{ens2021grand} define two grand challenges, i.e., \textit{supporting behavior with collaborators} and \textit{supporting cross-platform collaboration}, which highlight the necessity to facilitate collaboration across different levels of the virtuality continuum~\cite{milgram1995augmented}.
With the increasing maturity of various computing environments, particularly immersive ones, there are now greater opportunities for collaboration using diverse devices. 
This has garnered significant interest from researchers and has led to the emergence of asymmetric collaboration~\cite{grandi2019characterizing} and cross-device collaboration~\cite{brudy2019cross}. \added{Following this line of research, we aim to improve asymmetric collaboration with data visualization, especially in layout dimensionality.}

\subsection{Asymmetric Collaboration}
The wide array of devices across the augmented-virtuality spectrum has led to a wide spectrum of papers encompassing different types of immersive devices and their mediums of collaboration~\cite{frohler2022survey}.
Research has also developed a foundation for taxonomies to design applications allowing cross-device collaboration~\cite{brudy2019cross, olin2020designing}.

Much work has shown how design considerations could lead to better collaboration in Immersive Analytics.
Piumsomboon et al.~\cite{piumsomboon2019effects} discussed that when it came to the sharing of awareness cues across users in immersive collaborative settings that the sharing of the head-gaze was considered the most useful and easy to use.
From this, it was imperative that for cross-device collaboration to occur effectively, visual cues indicating user position would need to be implemented.
Müller et al.~\cite{muller2016virtual, muller2017remote} discusses how the usage of shared virtual objects aided in user discussion over physical objects.
To analyze so, they analyzed their communication behavior and defined their speech for spatial expressions into different categories such as ``Physical object'', ``Deictic speech'', ``Person'', etc.
We derived the analysis of communication based on these works with minor modifications that would align with the chosen task.

Many platforms have been developed for asymmetric collaboration, with platforms such as Spatial~\cite{Spatial}, VirBela~\cite{Virbela}, and Meta's Horizon Workrooms~\cite{MetaHorizonWorkrooms} showcasing the variety of asymmetric collaboration and the necessity of incorporating other devices into the design of immersive analytic tasks. 
However, these commercial platforms primarily emphasize social interactions, offering a consistent 3D environment for both PC and VR. In contrast, our focus lies in productivity scenarios, allowing us to explore a wider array of design options across these two environments' dimensionalities.
Kim et al.~\cite{kim2010hugin} introduced a framework enabling clients with various devices and immersion levels to interact within a common platform, specifically tabletop displays.
More recently, Saffo et al.~\cite{saffo2023through} argued that as the diversity of visualization types increases, a wider range of visualization metaphors becomes necessary to attain a higher degree of shared awareness. 
\added{In a related vein, Tong et al.~\cite{tong2023towards} conducted a controlled study to empirically evaluate the efficacy of asymmetric collaboration between immersive and non-immersive environments. 
They compared PC+VR asymmetric collaboration with PC+PC and VR+VR symmetric collaborations and discovered that asymmetric collaboration did not increase collaborative efforts, instead reducing the mental load associated with completing data analytic tasks.
Furthermore, they recommended optimizing asymmetric systems for the collaborators' respective devices to enhance the user experience.}

\added{However, their recommendation lacked empirical validation when juxtaposed with alternative asymmetric designs. 
For instance, instead of employing distinct designs, one could opt for a consistent design that prioritizes either PC or VR affordances. 
Thus, our investigation delves into an evaluation of various design alternatives in PC+VR asymmetric collaboration, particularly concerning layout dimensionalities.
}


\subsection{Layout}
\added{One of the benefits of immersive analytics is the unlimited space, allowing multiple windows to be viewed at the same time in different locations in the virtual space}\addedTwo{~\cite{liu2023datadancing}}
Layouting multiple windows in immersive analytics has been considered by researchers, and individual effectiveness could be altered as a result.
One earlier example of this is Shupp et al.'s~\cite{shupp2006evaluation} analysis of viewport sizes in Large High-Resolution Displays in which greater viewport curvature decreases performance times and are preferred by users.
More work has tried to analyze these benefits in immersive spaces, with works such as Ens et al.~\cite{ens2014personal} simulating this effect of multiple windows in an immersive, VR environment.
Lee et al.~\cite{lee2020shared} discusses that 2D visualizations are often paired by users using walls and 3D visualizations are positioned using the space around them, with other works such as \textit{Maps Around Me}~\cite{satriadi2020maps} defining different patterns such as spherical, spherical cap, and planar orientations. 
\added{Lisle et al.~\cite{lisle2020evaluating, lisle2021sensemaking} and Davidson et al.~\cite{davidson2022exploring} explore how people understand and apply multiple windows in immersive environments.
They discovered three of the main sensemaking arrangements that people employ in multi-window environments, semicircular, environment-based (those based on structures in the virtual environment), and planar.
As a result, we decided to use semicircular (3D) and planar (2D) arrangements of layouts, as these prior works showcase how people make sense of space in immersive environments. We excluded environment-based layouts given the lack of environment encodings possible on a 2D PC screen.
}

The decision towards how information should be displayed in immersive environments has been one that has been long considered among other works.
Liu et al.~\cite{liu2020design} discussed how given fewer multiples, a flat layout is preferred in cases where fewer multiples were present (4x3), in comparison to a larger set of multiples (12x3), where a semi-circular layout is preferred when interacting with 3D small multiples.
In Liu et al.'s future work~\cite{liu2022effects}, performance between a flat layout, a semicircular wrap-around layout, and a circular wrap-around layout were all relatively the same, despite participants preferring the semicircular wrap-around layout more than the other layouts.

\added{However, the effect of layout in asymmetric collaboration setting is still under-explored. We want to understand how layout and design between immersive and 2D desktop window management with} \addedThree{grid layouts} affect users' performance, collaboration, and preferences.


\section{Design and Implementation}
As outlined in the literature review, certain commercial social platforms offer accessibility from both PC and VR. However, these platforms exclusively offer 3D environments in both PC and VR settings. 
This work's aim is to systematically investigate design possibilities, specifically considering various combinations of dimensionalities, for PC+VR integration in a productivity context focused on data-driven decision-making.

\begin{figure*}
 \centering 
 \includegraphics[width=\textwidth]{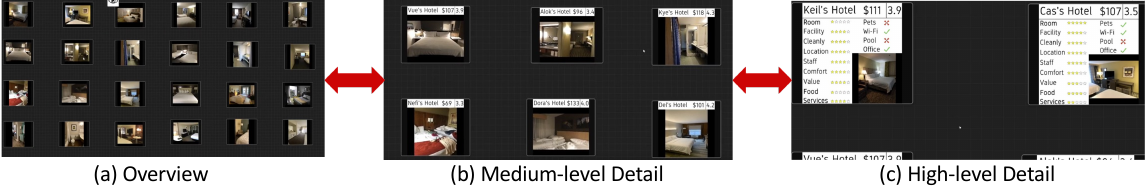}
 \caption{Demonstration of multiple-level of the hotel information and the amount of information presented on \cPCTwoD{} at each level.}
 \label{fig:pc-zooming}
\end{figure*}

\begin{figure}
 \centering 
 \includegraphics[width=\columnwidth]{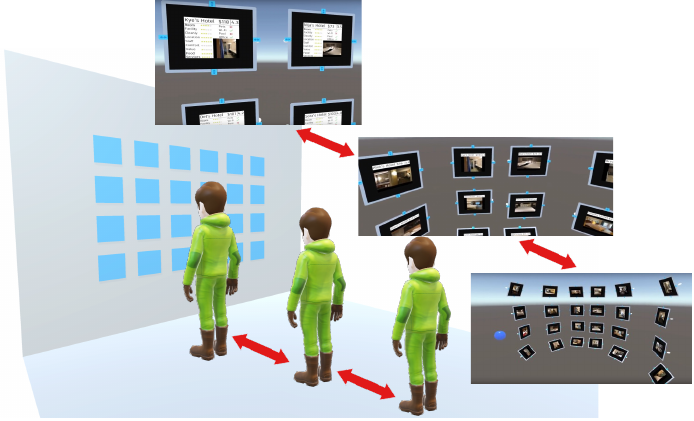}
 \caption{Illustration of bidirectional multiscale zooming in VR. Inspired by proxemic interaction~\cite{ballendat2010proxemic,badam2016supporting}, the presented level of detail is determined by the distance between the user and the views. Three levels of detail were provided in our study: far distance results in an overview with hotel images; medium distance adds hotel names, prices, and ratings; close distance further provides breakdown ratings and amenities. The same levels of detail were provided on PC, and users used the mouse scroll to switch between them.}
 \label{fig:vr-zooming}
\end{figure}

We chose to focus on a collaborative decision-making task from the literature~\cite{jetter2011materializing}, where collaborators are deciding on hotel choice based on various factors like price, rating, and amenities.
To allow users to collaboratively complete the task and based on previous work~\cite{tong2023towards,saffo2023through}, we proposed the following design goals:
\begin{itemize}
    \item \textbf{Facilitate hotel comparison.} The comparison of different hotels is a crucial aspect of the proposed task, and we aim to design user interfaces that effectively support this objective.
    \item \textbf{Support multi-scale navigation.} Viewing a large number of hotels' details simultaneously is impractical. The design approach focuses on enabling users to seamlessly switch between different levels of detail based on the specific information they need at any given time. 
    \item \textbf{Provide real-time visual awareness.} In collaborative settings, maintaining awareness of the collaborator's status, focus, and intent is crucial for effective communication and coordination. In line with this objective, we provided real-time visual awareness techniques to assist collaboration.
\end{itemize}

\subsection{Layouts in 2D and 3D}
\begin{figure*}
    \centering
    \includegraphics[scale=0.5]{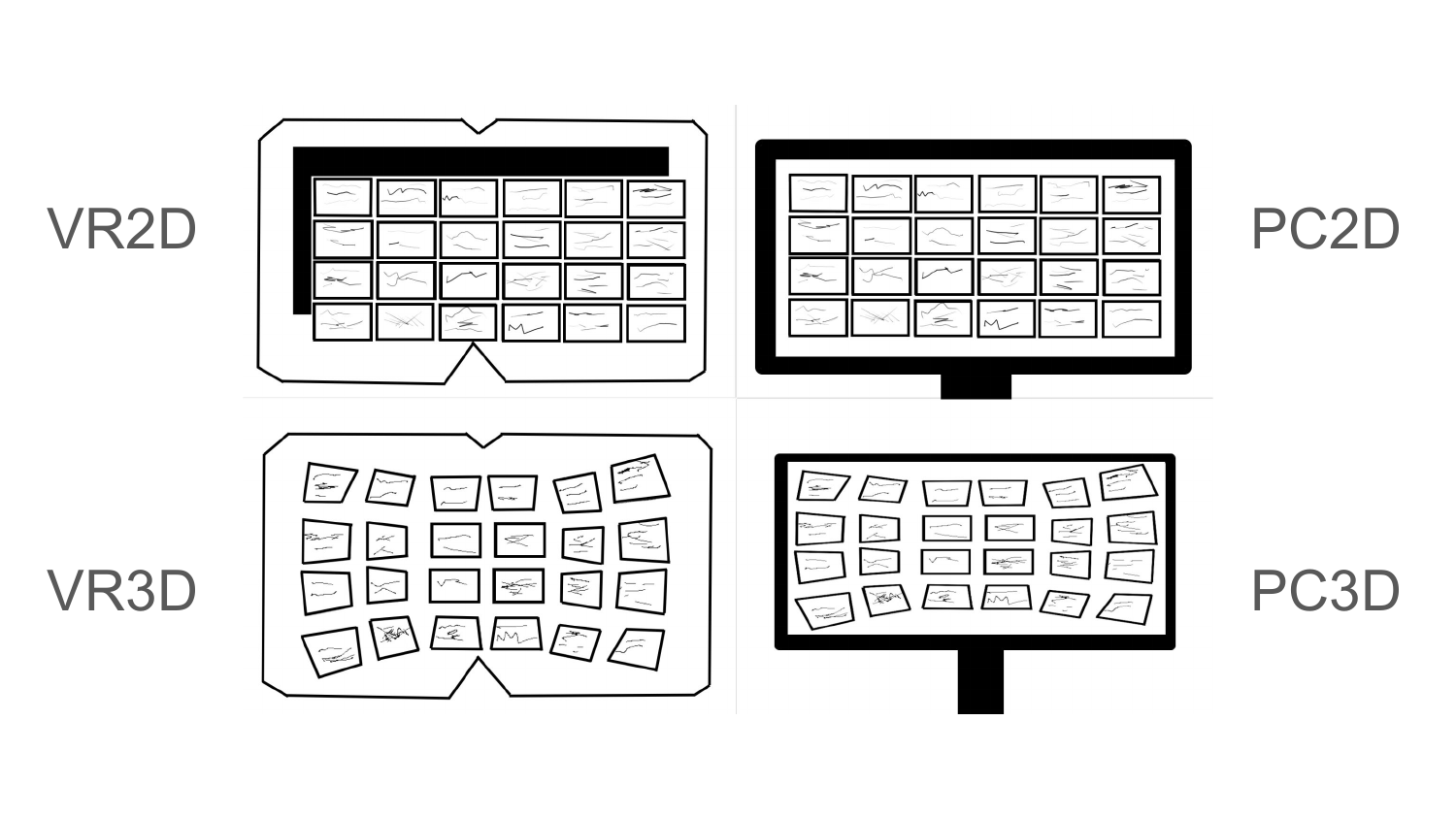}
    \caption{The four different views from the perspective of a user.}
    \label{fig:condition}
\end{figure*}

To enhance the process of \textit{hotel comparison}, we implemented \addedThree{a grid layout, akin to small multiples in data visualization} design in the experimental conditions. 
A small multiple refers to a series of visually similar content, such as graphs or charts, that enables easy comparison~\cite{tufte1985visual}.
In the task scenario, we utilized a \addedThree{design similar to small multiples} to represent a group of hotels, where each hotel's information was presented in a consistent format \addedThree{in the form of a panel or a window}. 
In a 2D environment, these panels are typically arranged in a grid format, as illustrated in \autoref{fig:condition}. 
In a 3D environment, the additional dimension allowed for different \addedThree{grid layouts}. 
As mentioned, Liu et al.~\cite{liu2020design,liu2022effects} examined various small multiple layouts in VR, including a flat layout (similar to a 2D environment), a semi-circular layout, and a fully circular layout. 
The semi-circular layout demonstrated overall benefits compared to other alternatives, and therefore, we selected it as the layout for the 3D environment.

To understand the collaborative aspects involving layout dimensionality, we decided to keep the information on the \addedThree{hotel panels} 2D, as factors related towards navigating a 3D visualization on a 2D display could add further complexity in identifying differences in layout.
2D visualizations are similar to the displays portrayed in practice by Meta Horizon Workrooms~\cite{MetaHorizonWorkrooms} and Virbela~\cite{Virbela}.

\subsection{Multi-scale Navigation}
Due to the extensive amount of information contained within the hotel comparison task, it is impractical to present every detail comprehensively.
This limitation arises from the restricted screen size of a PC, which cannot accommodate a large volume of information, and the fact that a VR user can only observe all hotel windows at a distance, with presented information being illegible.
To tackle this issue, we incorporated multi-scale navigation into the tested conditions. 
This means that the level of detail presented depends on the available display space. 
Essentially, by reducing the number of hotels displayed on a PC screen or within the VR user's field of view (FoV), we can increase the amount of detail provided for each individual hotel, and vice versa.
On a PC, users have the ability to utilize the mouse scroll to zoom in and out ~\addedThree{of the application window,  thereby increasing or decreasing the number of hotels they prefer to see on their screen and their corresponding sizes, with zooming out increasing the number of hotels and decreasing the size of each panel and zooming in decreasing the number of hotels and increasing the size of each panel as seen in} \autoref{fig:pc-zooming}.
\addedTwo{For PC, the three levels of detail were as follows: overview (100\%), medium detail (200\% of initial overview size), and high detail (400\% of initial overview size). }
In VR, users can physically move closer or farther away from the view, allowing them to determine the number of hotels that fit into their FoV, see \autoref{fig:vr-zooming}.
The VR multi-scale navigation was inspired by proxemic interaction~\cite{badam2016supporting,ballendat2010proxemic}, which focuses particularly on spatial relationship in this work, between people and the space around them and how it influences their interactions.
\addedTwo{For VR, the three levels of detail were as follows: initial view (outside of 2.5m), medium detail (within 2.5m), and high detail (within 1m). }

\begin{figure}
 \centering 
 \includegraphics[scale=0.5]{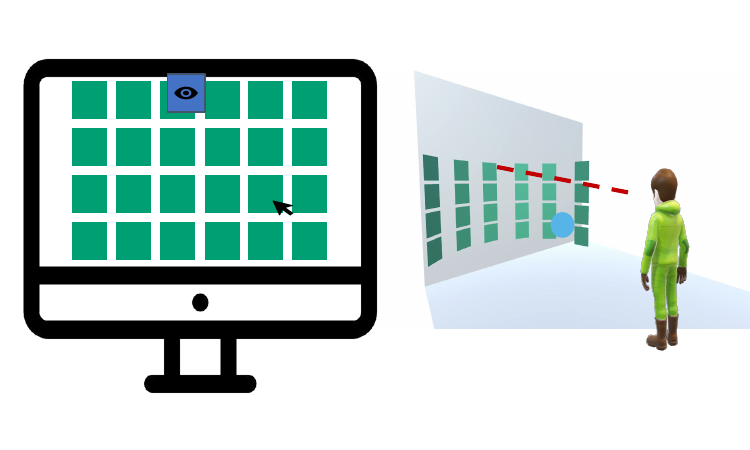}
 \caption{Demonstration of real-time awareness cues across PC and VR. Leveraging the depth-adaptive cursor technique~\cite{zhou2022depth,tong2023towards}, we are able to provide real-time awareness cues across platforms and dimensions (i.e., in \cPCTwoDVRThreeD{}). On PC (left), a moving indicator shows which window the VR collaborator is looking at. At the same time, in VR (right), an icon is rendered to indicate the PC collaborator's cursor position.}
 \label{fig:awareness}
\end{figure}

\begin{figure*}
    \centering
    \includegraphics[width=\linewidth]{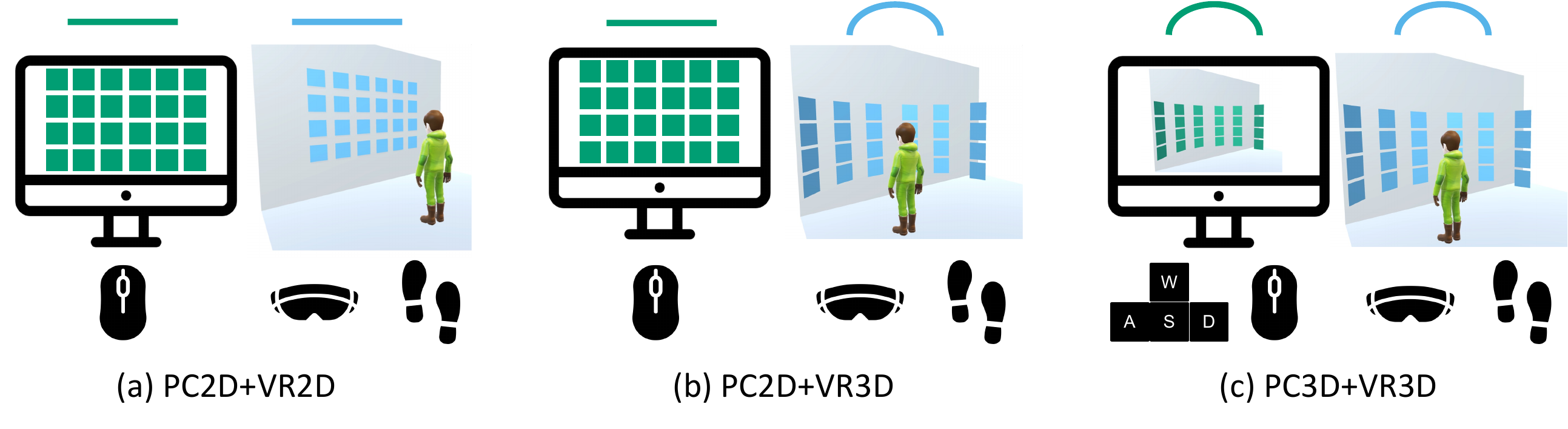}
    \caption{Three tested conditions in the user study. (a) \cPCTwoDVRTwoD{}, where the layout of views is 2D in both PC and VR; (b) \cPCTwoDVRThreeD{}, which involves a 2D layout for the PC collaborator and a curved 3D layout for the VR collaborator; and (c) \cPCThreeDVRThreeD{}, where both PC and VR have a 3D layout. 
    The PC collaborator uses pan\&zoom to navigate in 2D environments (i.e., \cPCTwoDVRTwoD{} and \cPCTwoDVRThreeD{}), while employing a combination of WASD keys and the mouse to navigate in 3D environments (i.e., \cPCThreeDVRThreeD{}). 
    This navigation method is similar to playing a first-person shooter (FPS) game and is commonly provided by commercial PC+VR social platforms. The VR collaborator walks in the space for both 2D and 3D layouts.}
    \label{fig:study-condition}
\end{figure*}

\subsection{Real-time Visual Awareness}
To enhance collaboration and communication, it is essential to provide users with contextual information about the content being referred to by their collaborators. To address this challenge, we have designed awareness cues that explicitly indicate the element under discussion.
Providing awareness cues across different virtuality is challenging, as both the dimensionalities and interaction modalities are different.
In order to bridge the gap between different virtuality levels, where dimensionalities and interaction modalities differ, we adopted the depth-adaptive cursor technique, building upon previous research~\cite{tong2023towards,zhou2022depth}. This technique allows us to project a 2D cursor into the 3D space and a 3D viewing direction into a 2D environment, as illustrated in \autoref{fig:awareness}. 
By leveraging this approach, users can easily navigate and interact with elements across different dimensionalities while maintaining a clear understanding of their collaborators' focus.
Additionally, we considered incorporating the view frustum into the conditions, as it has been found to be beneficial in VR/AR collaborative systems~\cite{piumsomboon2019effects}. 
However, based on recent studies involving PC+VR setups, it was discovered that providing the frustum may not be necessary~\cite{tong2023towards}, as the frustum might be unsuitable for the PC environment.

\subsection{Study Conditions and Implementation}
In the study, we specifically focused on investigating how the different dimensionalities in \addedThree{grid layouts} can impact collaboration in PC+VR settings. We conducted tests using three combinations: \cPCTwoDVRTwoD{}, \cPCTwoDVRThreeD{}, and \cPCThreeDVRThreeD{}, as depicted in \autoref{fig:study-condition}. 
Across all conditions, we provided consistent multi-scale navigation and real-time visual awareness functionalities.

In \cPCTwoD{}, we utilized SAGE3\footnote{\url{https://sage3.sagecommons.org/}} (Smart Amplified Group Environment), a successor of SAGE2~\cite{renambot2016sage2} for this implementation. 
It offers an infinite canvas as the working space, which is both zoomable and pannable, allowing users to overcome the limitations of a physical screen size. The concept of the infinite canvas has been adopted by various commercial tools, such as Miro and Google Jamboard. Within the SAGE3 workspace, we arranged the hotel information in a flat grid layout. 

In \cVRThreeD{}, we adopted a semi-circular layout to organize the hotel information. This layout takes advantage of the additional dimension offered by the VR environment, providing users with a curved arrangement of hotel details. The semi-circular layout offers improved visibility and facilitates easy comparison between different hotels~\cite{liu2020design,liu2022effects}. 
Conversely, in \cVRTwoD{}, we used a flat layout instead. 
In all VR conditions, users employed natural locomotion, which involves physically walking to navigate in the virtual environment.

In \cPCThreeD{}, we utilized the semi-circular layout in a 3D environment, similar to the one used in \cVRThreeD{}. 
To navigate within this 3D environment on a PC, the user needs to use the WASD keys on the keyboard for movement (forward, backward, left, and right) and the mouse for controlling the camera viewpoint. 
\addedThree{This input method for traversing in a 3D environment on PC is ubiquitous in gaming environments, with many 3D games such as the most popular game Minecraft and many commercial social PC-VR platforms, such as Spatial~\cite{Spatial}, VirBela~\cite{Virbela}, and Meta's Horizon Workrooms~\cite{MetaHorizonWorkrooms} using this input method for PC users.}

\added{We did not include \cPCTwoDPCTwoD{} and \cVRThreeDVRThreeD{} because these conditions are symmetric collaborative settings, i.e., collaborators use an identical device and digital tool for collaboration, and our focus is comparing the design alternatives in asymmetric collaborations. 
Additionally, these conditions were investigated by Tong et al.~\cite{tong2023towards} by comparing them to an asymmetric condition (similar to \cPCTwoDVRThreeD{}) to identify its potential benefits. 
We also excluded \cPCThreeDVRTwoD{} as this combination is likely to compromise both individual and collaborative effectiveness without evident benefits.
}

\section{User Study}

To understand the pros and cons of the three layout dimensionality combinations (\autoref{fig:study-condition}) in asymmetric collaborative decision-making, we designed and conducted a within-subject study.

\subsection{Task} \label{sec:Task}
The Hotel Search Task proposed by Jetter et al.~\cite{jetter2011materializing} was slightly modified in that participants were given a searchable set of 24 hotels, displayed in a fixed $6 \times 4$ layout, and were also given unlimited time to look for a hotel. 
\added{We did not want the time pressure to impact participants' behaviors.}
Participants were given a set of requirements that denoted their preferences when it came to which hotel to choose. 
These preferences took the form of a budget, overall rating, preferred amenities, or a preferred trait rating.
Each preference was also binary in whether or not it could or could not be met, with each preference having equal weight.
The PC participant was handed a list of these requirements on a piece of paper and the VR participant list existed in the virtual environment in a virtual canvas. 
This list differed across participants, training tasks, and conditions. 
The participants were not able to view the other person's preferences; however, they were allowed to communicate their preferences freely. 
The set of hotels and preferences was made to never allow for a hotel that fit all their preferences, forcing participants to compromise. 
Tie-breakers were broken by price.
This allowed for an optimal choice of hotel for each task. 

This task was chosen because it was a collaborative task that necessitated the use of communication, collaborative strategies, as well as spatial referencing.
Participants needed to communicate to describe preferences to one another.
Participants needed to devise a strategy to examine the data.
Participants needed to communicate location information when discussing hotels.
This task was chosen over other tasks such as ``Stegosaurus''~\cite{whiting2009vast} because of its relative simplicity, which allowed for straightforward derivation of collaborative actions. 
Other tasks might make it difficult to determine causality in collaborative behaviors, given the added complexity of the task.
Other tasks also run the risk of low collaborative effort, given that participants may feel less comfortable to collaborate if another participant is actively doing more analysis.

\subsection{Participants}
The study had 18 pairs of participants, totaling 36 participants \addedThree{who were recruited through a listserv of the university's student and faculty population.}
Participants were asked to come in pairs with someone they were familiar with to promote natural conversations and to alleviate the confounding factor of personal relationships affecting study results.
Participants were aged 19-30, with an average age of 24.47 years old and a standard deviation of 3.16.  
21 participants identified as male, 12 participants identified as female, 1 participant identified as non-binary, and 2 participants decided not to disclose their gender. 
Participants were assigned to be VR or PC users based on preference \addedThree{to prioritize familiarity with the VR/PC system or to give agency to participants when there was no preference.}
The participant would remain to the chosen device throughout the study for all testing conditions.
The VR users had an average VR experience rating of 2.94 out of 5 experience with a standard deviation of 1.43, with 1 having no experience and 5 having plentiful experience. 
All participants indicated that they had an average collaborative shared workspace experience (Such as Miro, Google Jamboard, or SAGE) of 2.25 out of 5, with a standard deviation of 1.44.

\subsection{Experimental Setup}
For VR, a Meta Quest Pro headset was used, providing $1800 \times 1920$ pixel resolution per eye and a 90Hz refresh rate, \addedThree{running through a SteamVR~\cite{steamvr} platform that was built on Unity~\cite{unity}}. 
The headset was wirelessly connected to a computer, \addedThree{the R10 Alienware Aurora Ryzen Edition PC}, with an AMD Ryzen 7 5800X 8-core processor and NVIDIA GeForce RTX 3080 graphics card, enabling free movement within the $4 \times 4 m$ space without cable impediments and leveraging the computer's powerful graphics processing. 
The windows were positioned $1 meter$ in front of the participant's starting position, each $0.6 \times 0.4 m$ in size\addedTwo{, with the second highest row of the windows being placed at eye level with each participant.}
This arrangement allowed participants to view all windows within their field of view. 
By walking through the space, participants could control how many windows were in their field of view at once, as well as the level of detail of the content (\autoref{fig:vr-zooming}).

For the PC setting, a 27-inch monitor with a $2560 \times 1440$ pixel resolution and 75Hz refresh rate was used \addedThree{to interface with the Desktop SAGE3 application~\cite{SAGE3}}. 
The monitor was connected to \addedThree{a separate computer with the same aforementioned model.}
Initially, all windows were displayed on the screen. 
Participants could scroll and zoom with the mouse to control how many windows fit on the screen at once, as well as the level of detail of the content (\autoref{fig:pc-zooming}).

All study conditions were able to be executed and interacted smoothly using the provided equipment and settings.
\addedThree{To emulate a remote working environment without creating or interfacing with a voice communication application, participants would be located in the same physical environment with temporary walls to block their vision to the other participant while still allowing the other participant to hear their voice. }

\subsection{Study Design and Procedure}
The user study followed a full-factorial within-subjects design, with conditions balanced using a Complete Latin square to minimize the order effect. 
\addedTwo{Participants would remain as the VR user or the PC user for the purposes of time and to mirror collaborative actions on other asymmetrical applications.}
The study lasted for a total of around 90 minutes on average. 
Participants were initially welcomed and reviewed a consent form \addedThree{and would be informed of potential VR sickness or other discomfort due to the application and were welcome to take extra breaks as needed.}
Then, we briefly introduced the study’s objectives and procedural steps. 
\addedThree{To provide a realistic application scenario, we instructed participants to complete the task as quickly and as accurately as they could.}
\addedThree{All data from the experiment and the recording would be anonymized by the researcher conducting the experiment.}
Following this introduction, participants proceeded to the various components of the study as follows:

\noindent\textbf{Preparation}: 
We asked participants to adjust the chair height to a comfortable level for PC and adjust the Quest Pro headset for VR before they started. 
We confirmed that all participants were in comfortable conditions and could see the text in all environments clearly.

\noindent\textbf{Main Task}: We followed the procedure below for each condition.

\textit{Training}:
Participants were asked to complete a simple version of the Hotel Search Task (Section \ref{sec:Task}) with a smaller data set ($4 \times 3$ hotels) to get familiar with the new collaboration environment.
Participants were free to inquire about interactions or tasks. The training concluded once participants were proficient with tasks and especially the interactions, generally taking 3-5 minutes.

\textit{Study Task}:
Upon completion of the training session, participants proceeded to the study task. 
The study task was the same as the training task but with a different and larger data set ($6 \times 4$ hotels).
Participants had no time limit for task completion but were encouraged to prioritize accuracy and efficiency. For the VR environment, we reset the participants’ position to the center of the room and had them face the same initial direction before each study task started.
\addedTwo{Participants would indicate they have completed the task by noting to the researcher that they have finished the task. The time would be recorded and the participants would then tell the researcher their final answer.}

\textit{Break}:
Participants were given a mandatory 5-minute break to prevent stress, burden and to relieve potential physical demands of VR.

\noindent\textbf{Ending}:
After the completion of all three tasks, the participants \addedTwo{each separately} filled out questionnaires for the evaluation and ranking of all conditions \addedTwo{on a Google Form}, followed by a semi-structured, recorded interview.


\subsection{Measures}
The following measures are gathered from the perspective of \textit{effectiveness}, \textit{collaborative effects}, \textit{group awareness}, and \textit{preferences}.

\textbf{Effectiveness.}
\textit{Completion time} was measured from the moment the operator said to begin until the users indicated they were finished.
\textit{Choice accuracy} was also measured, given that each hotel had an individual count of compromises and issues to resolve ties.
\textit{Task load} and \textit{collaborative engagement} were also evaluated with the 7-point Likert scale NASA TLX questionnaire~\cite{sandra2006nasatlx}.

\textbf{Collaborative Effects.}
\textit{Communication effectiveness, strategy, and coordination} were measured through a quantitative assessment of words spoken. 
To do this, two independent coders coded the transcripts of the audio recording.
Each coder coded 12 sessions, with six overlapping sessions.
The six overlapping sessions allowed us to evaluate the inter-coder reliability using Cohen's Kappa of $>$ 0.7.
The coding scheme was derived from Mahyar and Tory~\cite{mahyar2014supporting} and was modified to fit with the task specified by this work.
The coding scheme with examples is shown in \autoref{tab:code-table}. Other statements that are relevant but not categorized into these six categories become uncategorized.
Additionally, we were particularly interested in \textit{spatial referencing} behaviors, as we anticipated it would be an important process in collaborative work.
To measure spatial referencing, we utilized a coding scheme similar to Müller et al.~\cite{muller2017remote} in which types of spatial expressions were grouped and the frequency was denoted for each individual. 
This coding scheme is shown in \autoref{tab:spatial-code-table}.
\added{Lastly, we were interested in seeing whether different layouts affect memorability during collaboration. Therefore, we counted the number of times when users mentioned that they forgot the position of the information, such as, ``\textit{Where is that?}'' and ``\textit{Do you remember ...?}'', as \textit{forget}. We also counted the number of times when another participant directly responded to one of the questions asked above, such as ``\textit{So I remember the one is also three.}'' as \textit{recall}.}

\textbf{Group Awareness.} To measure group awareness, we utilized the 7-point Likert scale behavior engagement questionnaire from Networked Minds Measure of Social Presence~\cite{biocca2001networked}. 

\added{\textbf{Preference.} To assess preference, we requested participants to rank the conditions. Additionally, we gathered detailed rankings on user-\textit{friendliness}, \textit{productivity}, and \textit{communication ease}.}

\begin{table*}[]
\centering
\caption{The coding schema of the transcription for \textit{Collaborative Effects}, derived from \cite{mahyar2014supporting}.}
\label{tab:code-table}
\resizebox{\textwidth}{!}{%
\begin{tabular}{|l|l|l|l|}
\hline
\textbf{Category} & \textbf{Code} & \textbf{Definition} & \textbf{Example} \\ \hline
Discuss Hypothesis & DH & A statement of noting a discussion of a hypothesis, either a claim or a comparison & ``I think John's is the best.'' \\ \hline
Strategy Coordination & SC & A statement outlining coordinating strategies of what the pair should perform next & ``I get the left half, and you get the right half.'' \\ \hline
Personal Information & PI & Sharing information pertinent that only they would know & ``I need a 4.2 overall. It needs to have a pool.'' \\ \hline
Seeking Awareness & SA & Questioning and actively looking for knowledge of a location & ``Where was John's hotel?'' \\ \hline
Verbalize Findings & VF & Communicating a new piece of knowledge based on a recent finding of information & ``John's hotel has three compromises for me.'' \\ \hline
Question Findings & QF & Questioning the findings & ``How many compromises does this have for you?'' \\ \hline
\end{tabular}%
}
\end{table*}

\begin{table*}[]
\centering
\caption{The coding schema of the transcription for \textit{Spatial Referencing}, derived from \cite{muller2017remote}.}
\label{tab:spatial-code-table}
\resizebox{0.7\textwidth}{!}{%
\begin{tabular}{|l|l|l|}
\hline
\textbf{Category} & \textbf{Definition} & \textbf{Example} \\ \hline
Exact Position & A statement noting a coordinate position of a panel & ``It's at Column 3, row 2.'' \\ \hline
Relative Position & A statement noting the position of a panel relative to themself  & ``It's to the right.'' \\ \hline
Mix & A combination of the exact position and relative position & ``Third from the right.'' \\ \hline
Deictic Speech & A statement that does not convey location information & ``It's over there.'' \\ \hline
\end{tabular}%
}
\end{table*}

\subsection{Hypotheses}
\addedThree{Four} hypotheses have been made in order to assess the differences PC users and VR users have in asymmetric collaboration.

\textbf{Effectiveness.} We believe that \cPCTwoDVRThreeD{} will have the best individual effectiveness ($H_{eff}$), given that the environment is suited to both meet the needs of the PC and VR users. Specifically, we have the following three sub-hypothesis.
\begin{enumerate}[leftmargin=*,topsep=0pt,label=\textcolor{blue}{(\Alph*)}]
    \item $H_{effA}$: users in \cPCTwoDVRThreeD{} completed the task the fastest.
    \item $H_{effB}$: the accuracy of the task should remain similar since we did not provide additional functionalities for specific conditions.
    \item $H_{effC}$: users should perceive the least task load in \cPCTwoDVRThreeD{}.
\end{enumerate}


%
\textbf{Collaborative Effects.} We believe that \cPCTwoDVRTwoD{} will have better collaborative effectiveness ($H_{coll}$), given that in the system, the VR user's environment appears extremely similar to the PC user's environment. 
Specifically, we have the following three sub-hypothesis. 
\begin{enumerate}[leftmargin=*,topsep=0pt, label=\textcolor{blue}{(\Alph*)}]
    \item $H_{collA}$: participants in \cPCTwoDVRTwoD{} should have the fewest instances of spatial references.
    \item $H_{collB}$: participants in \cPCTwoDVRTwoD{} should have the most instances of discussing the hypothesis.
    \item $H_{collC}$: participants in \cPCTwoDVRThreeD{} and \cPCThreeDVRThreeD{} should have more instances of recall than \cPCTwoDVRTwoD{}.
\end{enumerate}

\textbf{Group Awareness.} We derive that given the ``eyes-and-shoes'' principle defined by Saffo et al.~\cite{saffo2023through}, we can expect that given the diverse extreme of the virtuality continuum that the users' experience, greater group awareness will be required. As a result, \cPCTwoDVRThreeD{} should have the most social behavior ($H_{aware}$).

\added{\textbf{Preferences.} We expect that \cPCTwoDVRThreeD{} will rank the best as both individuals are able to use the most comfortable representation for the working device~\cite{tong2023towards} ($H_{preference}$). Specifically, \cPCTwoDVRThreeD{} will be preferred in terms of user-friendliness and productivity but not communication ease as asymmetry will lead to harder spatial referencing.}

\begin{figure*}
 \centering 
 \includegraphics[width=1.00\textwidth]{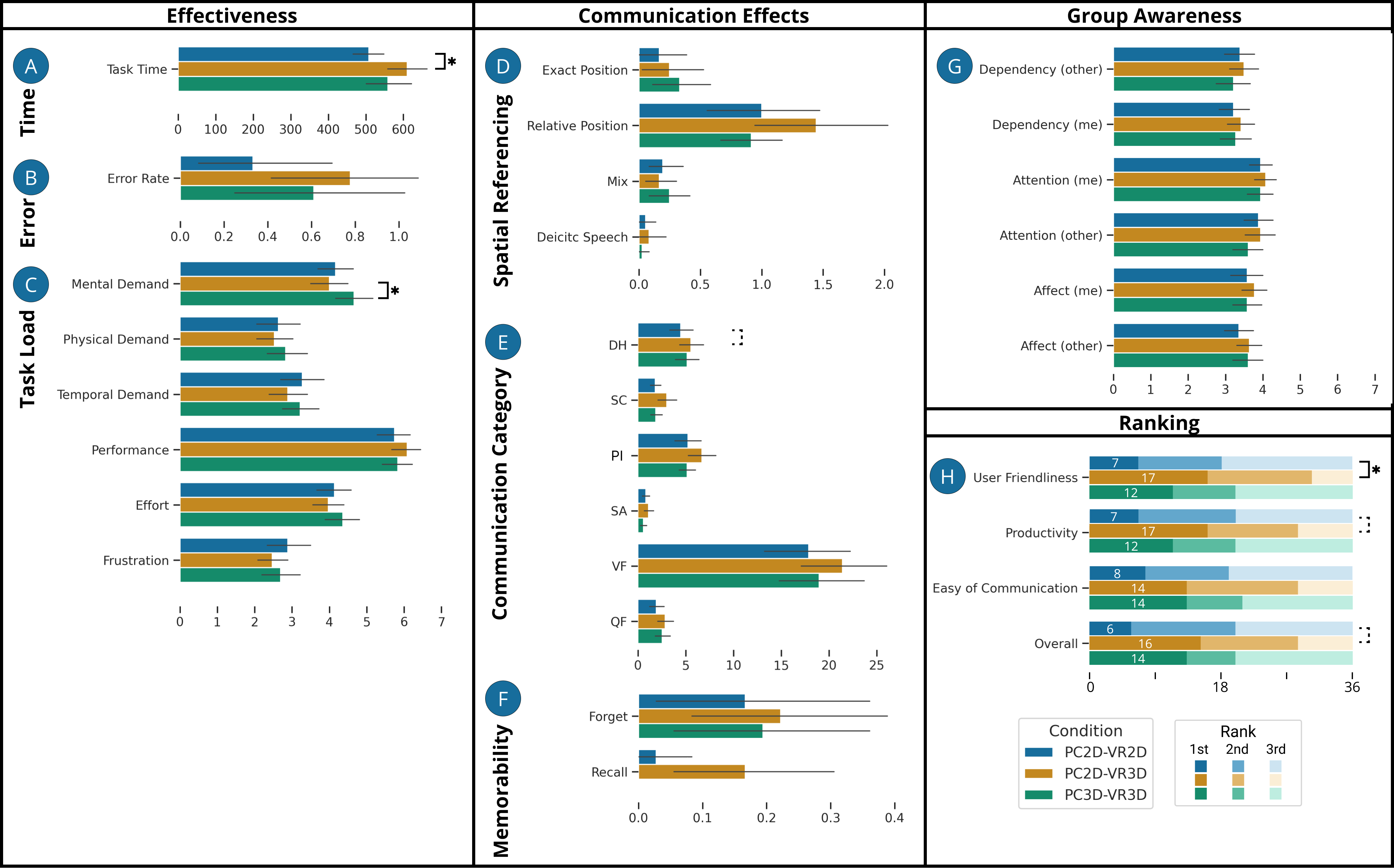}
 \caption{The figure shows the mean and 95\% confidence intervals across all three conditions with the result of all measures, subdivided by relevant hypotheses. (a) task time in seconds, (b) error rate, with 0 being the correct choice, 1.0 being the second best choice, 2.0 being the third best, etc., (c) task load, (d) spatial reference frequency, (e) communication category frequency, (f) memorability frequency, and (g) group awareness. Additionally, a stacked bar chart (h) shows the distribution of ranking. Dotted lines indicate marginal significance, and solid lines with stars indicate the level of significance, with symbols denoting a p-value of less than .05 ($*$), .01 ($**$), and .001 ($***$).}
 \label{fig:all}
\end{figure*}

\section{Results}
We applied different tests and analytics methods to analyze the data.
For completion time, we first applied a log transformation to meet the normality assumption and then used \textit{repeated-measure ANOVA} with \textit{Tukey post-hoc} analysis.
For all other quantitative measures, we conducted \textit{Friedman tests} with \textit{Nemenyi post-hoc} analysis to test if there are significant differences.
For qualitative feedback, we adapted affinity diagramming~\cite{hartson2012ux} to analyze the \added{subjective feedback of the individuals from the} transcribed interview recordings.
Significance values are reported for $p \le .1(\cdot)$, $p < .05(*)$, $p < .01(**)$, and $p < .001(***)$.
The results can be seen in Figure ~\ref{fig:all}.


\subsection{Effectiveness}
For \textit{completion time}, we found there is a significant difference in completion time ($F=4.64,$\pValue{0.01}, $\ast$). \autoref{fig:all}A showed that participants in \cPCTwoDVRTwoD{} completed the task significantly faster than \cPCTwoDVRThreeD{} (\pValue{0.02}, $\ast$). \textbf{$H_{effA}$ is rejected.}
For \textit{accuracy} (\autoref{fig:all}B), there were no significant differences between the three conditions and as a result, \textbf{$H_{effB}$ is accepted.} 
With \textit{task load} (\autoref{fig:all}C), there were no notable differences except for mental demand (\fried{6.34}{0.04}, $\ast$). The study revealed that \cPCThreeDVRThreeD{} leads to considerably higher mental demand compared to \cPCTwoDVRThreeD{} (\pValue{0.04}, $\ast$). \textbf{$H_{effC}$ is partially accepted.} 

As a result, we \textbf{partially rejected $H_{eff}$} given \cPCTwoDVRThreeD{} did not provide the best effectiveness but could reduce the mental demand compared to \cPCThreeDVRThreeD{}.

\subsection{Collaborative Effects}
For \textit{spatial referencing} (\autoref{fig:all}D), there is no significant difference between the three conditions. \textbf{$H_{collA}$ is rejected.}
For the number of instances in \textit{communication categories} (\autoref{fig:all}E), we have observed a marginal variance in the number of DH (discuss hypothesis) (\fried{5.81}{0.05}, $\cdot$) instances between the three conditions. More specifically, we found that \cPCTwoDVRThreeD{} exhibits a marginal significantly higher number of DH instances compared to \cPCTwoDVRTwoD{} (\pValue{0.09}, $\cdot$). \textbf{$H_{collB}$ is rejected.}
For \textit{memorability} (\autoref{fig:all}F, there is a notable variance in recall between the three conditions based on the outcome (\fried{8.85}{0.01}, $\ast$). However, we cannot identify any significant differences when comparing each pair of conditions. Therefore, \textbf{we cannot accept $H_{collC}$}. As a result, we \textbf{rejected $H_{coll}$} given \cPCTwoDVRTwoD{} did not have the best communication effects, yet \cPCTwoDVRThreeD{} motivated more discussion.

\subsection{Group Awareness}
Based on the collected data shown in \autoref{fig:all}G, we did not find a significant difference between the three conditions in terms of \textit{social presence}. Thus, we \textbf{reject $H_{aware}$} that \cPCTwoDVRThreeD{} required more group awareness.

\subsection{Preferences Rankings.} In general, participants rated \cPCTwoDVRThreeD{} better than \cPCTwoDVRTwoD{}, with marginal significance (\pValue{0.06}, $\cdot$). Specifically, in \autoref{fig:all}H, \cPCTwoDVRThreeD{} was considered significantly more user-friendly than \cPCTwoDVRTwoD{} (\pValue{0.02}, $\ast$). Additionally, it was marginally rated as more productive than \cPCTwoDVRTwoD{} (\pValue{0.06}, $\cdot$). Therefore, $H_{preference}$ \textbf{is partially accepted }.
We were also interested in whether the perceived preference changed for different dimensionalities in the same environment and did not find statistical significance in the analysis.


\subsection{Qualitative Feedback}
In \cPCTwoDVRTwoD{}, participants with the 2D PC commented they are ``comfortable'' and ``easy to interact with'' in this environment. Generally, participants dislike the flat layout and ``more physical movement'' in VR 2D, however, since both of the users are using the interface with the same dimensionality, VR P11 mentioned that ``I mean the one that I preferred the most was probably the linear layout. Um, why is because the way that we executed our strategy, you know, I didn't need curvature, right?''. 

For \cPCTwoDVRThreeD{}, participants are able to use their best-fit environment with the devices, it makes the collaboration with ``better communication''. For example, VR P7 commented that ``I mean the second one was easier just because I know he [PC P7] was having trouble with the third one. So it just like made communication easier.'' Moreover, compared to \cPCTwoDVRTwoD{}, the VR users are happy by reducing their physical movement. VR P13 pointed out that ``overall the third one. Cause it was just like the easiest one to look at. ... the third one [\cPCTwoDVRTwoD{}] was the most like circular, so I just kind of had to move my head.''

Lastly, for \cPCThreeDVRThreeD{}, by introducing PC 3D, again, we made both users collaborate in the same dimensionality. However, participants dislike the interaction because it requires bimanual input, which provides a steep learning curve. For example, PC P15 stated that ``[...] because I used to play video games like a very long time ago, [...] it took me a while to navigate things with. So, but if someone, like someone else already knew how to play video games, it would be easier for them to navigate.''

\section{Key Findings and Discussions} 

\textbf{Individual effectiveness significantly influenced user preference: \cPCTwoDVRThreeD{} was overall preferred.}
We consider 2D as the optimal environment for our tested task on a PC because participants frequently complained about using a mouse and keyboard to navigate a 3D environment.
Meanwhile, leveraging the 3D display space was found to be preferred over having a flat layout by Liu et al.~\cite{liu2020design,liu2022effects}.
Based on their findings, the 3D environment was more ideal for the collaborative task for VR users, \addedThree{given the large amounts of information panels}.

Overall, our collected data found that participants favored conditions in which individual effectiveness was the priority. 
This effect was predominantly seen in the \cPCTwoDVRThreeD{} condition, where both the PC user and the VR user had the interface be individually effective to both users.
This was felt in multiple participant responses, in which \cPCTwoDVRThreeD{} significantly ranked higher than \cPCTwoDVRTwoD{}, such as in overall preference, productivity, and user-friendliness.
For overall preference, when adjusted according to user type, the VR users preferred the \cPCTwoDVRThreeD{} condition more than the \cPCTwoDVRTwoD{} condition, which aligns well with prior study from Liu et al.~\cite{liu2022effects}.
\addedThree{This outcome means that designers should utilize the full extent of an individual's environment when designing in terms of optimizing user experience. }

\vspace{4pt}\noindent\textbf{Collaborative effectiveness was positively correlated with the completion time: \cPCTwoDVRTwoD{} was overall fastest.}
Despite the fact that \cPCTwoDVRThreeD{} was the most preferred condition, \cPCTwoDVRTwoD{} performed significantly faster with respect to task completion time than \cPCTwoDVRThreeD{}.
A possible reason why \cPCTwoDVRTwoD{} significantly outperformed \cPCTwoDVRThreeD{} is because of the added collaborative effectiveness in the VR user and the PC user sharing the same view.
A reason why \cPCThreeDVRThreeD{} did not have the same level of performance despite this added collaborative effectiveness could be caused by the added mental demand required to operate the system for the PC users, hindering the individual effectiveness considerably and therefore making the condition result in lower completion time.

A possible design implication is that VR users could find it easier to adjust to a flat display in contrast to PC users would with a 3D display. This is because VR users are more capable of compromising individual effectiveness given the improvement provided to collaborative effectiveness.


\vspace{4pt}\noindent\textbf{Participants did not perceive an obvious difference when the layout dimensionality of their collaborator was changed.}
Many participants noted similarities between conditions.
This was the intention of the study design, given that in \cPCTwoDVRTwoD{} and \cPCTwoDVRThreeD{}, the PC user was given the same display, and in \cPCTwoDVRThreeD{} and \cPCThreeDVRThreeD{} the VR user was given the same display.
PC P9 wrote ``I don't remember any significant moments where things felt different in both the 2D ones.'', VR P4 wrote ``there wasn't much of a noticeable difference between the first and second experiments because I felt like really nothing changed'' and VR P7 wrote I couldn't really notice that my partner had a completely different experience.''
Other participants did, however, notice this difference in collaboration, with PC P6 claiming ``When my partner had 2D he went too fast, and I had the most problem keeping up with him'' and PC P2 saying ``The 2nd and 3rd condition was essentially the same for me because PC was 2D, but I ranked the VR3D higher because we were a lot more efficient'', however, this was a small minority given most participants did not notice a collaborative difference across conditions.
Despite users not noticing the visual difference in the environment, collaborative effects were still felt as participants performed differently given the collaborative effort required by them or their partners.
This is indicated by \cPCTwoDVRTwoD{} having a lower completion time than \cPCTwoDVRThreeD{}. 
\addedThree{Because participants were not able to perceive a difference with collaboration despite change in the others' environment, designers are able to change a user's environment without affecting the other user as long as collaboration remains similar for both users. 
This insight is important for researchers as collaboration may appear similar for users despite changes in their collaborators environments. Researchers may consider informing the user of their collaborator's environment, given that the user may not be able to notice otherwise.}

\vspace{4pt}\noindent\textbf{Having consistent dimensionality across PC and VR resulted in a ``follow the leader'' workflow, while different dimensionalities (\cPCTwoDVRThreeD{}) promoted hypothesis discussion and led to more collaborative effects.}
``Follow the leader'' is a workflow in asymmetric collaboration identified by Saffo et al.'s~\cite{saffo2023through}, where users with views designated to their task took on a leadership position. 
We observed that the PC user in the \cPCTwoDVRTwoD{} user was more frequently the \textit{leader} given that the view for the VR user was altered to correspond to the PC user, making the PC user the leader, with a similar effect happening with the VR user being a leader in the \cPCThreeDVRThreeD{} condition.
We found that this ``follow the leader'' may have occurred, given in the communication analysis, the condition with the most DH and SA in the pairwise comparison was \cPCTwoDVRThreeD{}, the condition in which individual effectiveness was prioritized for both users.
In \cPCTwoDVRThreeD{}, since the participants are \textit{equal} in terms of leadership, they both have equal opportunities to DH, as DH could be seen as a type of \textit{leadership} communication, given the leader discussed hypotheses necessary to the completion of the trial and if both participants felt as if neither was \textit{in charge}, both shared equally.
Similarly, SA could be seen as a \textit{follower} communication type, being inquisitive. Given that both participants may be seen as equal, both participants could feel inclined to ask questions to the other.
The analysis does not find strong, concrete evidence of ``follow the leader'' attributes, but these two collaborative effects in communication could have been from ``follow the leader''.

One conclusion that could be made clear as a result of \cPCTwoDVRThreeD{} having the most DH and SA occur could be that the inconsistency in layout dimensionalities leads to more discussion.
Given that users are in an environment where individual effectiveness is prioritized, both users could feel equally inclined to discuss amongst one another, thus leading to more DH and SA than the other conditions. 
\addedThree{Designers can leverage the ``follow the leader'' workflow to allow for the user in an individually effective environment to be the \textit{leader} or design an equally effective environment to not produce these effects.}

\vspace{4pt}\noindent\textbf{Navigating a 3D environment was cumbersome on a PC: \cPCThreeD{} was frequently complained about and \cPCThreeDVRThreeD{} was more mentally demanding.}
Specifically for PC users, the \cPCThreeDVRThreeD{} condition was found to have higher mental demand than the \cPCTwoDVRThreeD{} users.
Possible reasons for this discrepancy could be because of the extra degrees of freedom and extra controls required to interact with the 3D environment.
Users moving with one hand and looking at the environment with their mouse, combined with general unfamiliarity with 3D environments on a PC screen, could contribute to the added mental demand perceived by the users. 
As a result, the effort required to manipulate the camera and understand the environment contributed to this effect of \cPCThreeDVRThreeD{} users having more added mental demand.
This claim was also facilitated by some of the users, with PC P16 saying about the \cPCThreeDVRThreeD{} condition ``Using both hands and communicating with my partner at the same time was very challenging.'' and PC P3 claiming ``I found the use of the keyboard [...] a little irritating and took time to move around. So I was a bit slower, so the partner was initially taking the lead. Once I got used to it, I could pace up with my partner and started to take the lead with the preferences said first from my end".

\vspace{4pt}\noindent\textbf{Different layout dimensionality combinations resulted in similar spatial referencing behavior.} 
Although we identified influences of layout dimensionality combinations for asymmetric collaboration, no statistical significance was found between spatial referencing types and amounts across conditions. 
This suggests users employed similar methods to discuss items spatially, regardless of dimensionality differences. 

This result contrasts our original expectation that the distinct visual presentation and navigation affordances across conditions would confer different levels of spatial awareness. 
Specifically, the clear difference between 2D and 3D on PC should manifest in the data. We posit the designed visual awareness tool (\autoref{fig:awareness}) reduced the required explicit spatial referencing (i.e. verbal communication), with participants instead checking the visual indicator implicitly. 
While we still consider the total spatial referencing action count affected by dimensionality combinations, future studies should collect data on implicit spatial referencing actions (e.g., looking at the visual indicators) to verify this.

\addedThree{This indicates that spatial referencing behavior is preserved despite differences in layouts, given that spatial linearity is preserved. Designers wanting to allow users to collaborate across dimensions should preserve relative spatial positions to allow for seamless collaboration in regards to referencing behaviors.}

\section{Generalization, Limitation and Future Work}


Our study tested different combinations of dimensionalities for PC+VR asymmetric collaborative decision-making. 
In order to gain a nuanced understanding of PC+VR collaboration, we adapted a well-established hotel search task~\cite{jetter2011materializing}. 
Given the nature of this task, we employed a \addedThree{grid layout} design, where all windows appeared identical except for displaying distinct hotel information. 
In more complex data-driven decision-making scenarios, collaborators may require access to diverse information presented in different visual formats, such as a dashboard. Although we believe that many of our findings could be applied to such scenarios, it is important to note that due to differences in navigation behaviors, we might observe contrasting collaborative actions, especially concerning spatial referencing. Hence, further investigation through future studies is necessary.

Additionally, it is worth mentioning that although the \addedThree{grid layouts} were 3D in certain conditions, the visual content within each individual window remained inherently 2D. As a result, our future direction involves examining task scenarios that involve 3D content or a combination of 2D and 3D content. This will allow us to explore the impact of spatial depth and visual immersion on collaborative decision-making in PC+VR environments more comprehensively.
\addedThree{Similarly, the investigation of different layouts such as those with larger sizes or those that are non-linear in their placement may see alternate effects in regards to spatial referencing behavior or collaborative effects given that the relative spatial positioning may be affected given the added depth axis when contrasting 2D and 3D environments. Designs would need to accommodate for potential occlusion given the added depth axis may obstruct windows and would need to be evaluated in future work.}

We ensured that our participants had the necessary interactions to successfully accomplish the study task. Nevertheless, we acknowledge the potential benefits of incorporating additional interactions, such as the ability to move and resize windows. 
By introducing these interactions, we would face new technical challenges, including real-time layout synchronization between 2D and 3D environments. However, such additions would also enhance collaborative behaviors, for instance, allowing for the interactive formation of clusters~\cite{luo2022should,davidson2022exploring}. 
Furthermore, as more interactions are introduced, awareness cues will play an even more crucial role, given the increased activity of both users and visual content. Therefore, improving awareness techniques to effectively support more interactive experiences is imperative.

Lastly, our study revealed that employing both 2D and 3D environments within VR presents its own set of advantages and disadvantages. One potential solution is to offer VR users the option to switch between 2D and 3D environments or even implement automatic switching. For instance, using 3D for individual work and transitioning to 2D for collaborative discussions could be beneficial. This approach would allow users to leverage the strengths of both environments based on the specific task or stage of the collaborative process.



\section{Conclusion}

This paper presents an empirical study that utilizes both quantitative and qualitative analysis to examine the impact of different designs of asymmetric collaborative systems between PC and VR on the collaborative decision-making process. The study focuses on a classic data-driven collaborative task, specifically the hotel search task, and employs a \addedThree{grid layout} design to present the hotel information. The \addedThree{grid layout} varies in dimensionalities across different conditions, namely \cPCTwoDVRTwoD{}, \cPCTwoDVRThreeD{}, and \cPCThreeDVRThreeD{}.
To facilitate task completion, we incorporate semantic zooming and awareness support in both environments.
Our findings indicate that optimizing the individual environment enhances satisfaction and improves engagement. On the other hand, interfaces that prioritize reducing collaborative cost lead to faster completion times. However, we observed that navigating a 3D environment on a PC was ineffective for our tested task, although no similar trend was observed in VR.
We believe that our results contribute to an empirical understanding of the collaborative experience in asymmetric collaboration and can serve as inspiration for future designs.



\bibliographystyle{ACM-Reference-Format}
\bibliography{references}


\end{document}